# Concept-oriented model:
# Modeling and processing data using functions

Alexandr Savinov

http://conceptoriented.org

17.11.2019

**ABSTRACT**

We describe a new logical data model, called the concept-oriented model (COM). It uses mathematical functions as first-class constructs for data representation and data processing as opposed to using exclusively sets in conventional set-oriented models. Functions and function composition are used as primary semantic units for describing data connectivity instead of relations and relation composition (join), respectively. Grouping and aggregation are also performed by using (accumulate) functions providing an alternative to group-by and reduce operations. This model was implemented in an open source data processing toolkit examples of which are used to illustrate the model and its operations. The main benefit of this model is that typical data processing tasks become simpler and more natural when using functions in comparison to adopting sets and set operations.

## 1 Introduction

### 1.1 Who Is to Blame?

Most of the currently existing data models, query languages and data processing frameworks including SQL and MapReduce use mathematical *sets* for data representation and *set operations* for data transformations. They describe a data processing task as a graph of operations with sets. Deriving new data means producing new sets from existing sets where sets can be implemented as relational tables, collections, key-value maps, data frames or similar structures.

However, many conventional data processing patterns describe a data processing task as deriving new *properties* rather than sets where properties can be implemented as columns, attributes, fields or similar constructs. If properties are represented via mathematical *functions* then this means that they are main units of data representation and transformation. Below we describe several typical tasks and show that solving them by means of set operations is a problem-solution mismatch, which makes data modeling and data processing less natural, more complex and error prone.

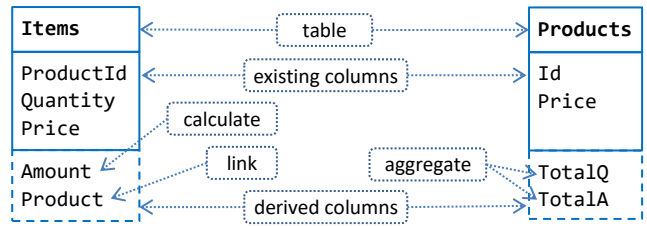

**Figure 1: Example data model**

*Calculated attributes.* Assume that there is a table with order `Items` characterized by `Quantity` and `Price` attributes (Fig. 1, left). The task is to compute a new attribute `Amount` as their arithmetic product. A solution in SQL is almost obvious:

```
SELECT *, Quantity * Price AS Amount          (1)
FROM Items
```

Although this standard solution seems very natural and almost trivial, it does have one subtle flaw: the task was to compute a new *attribute* while this query produces a new *table*. Then the question is why not to do exactly what has been requested by producing a new attribute? Why is it necessary to produce a new table (with a new attribute) if we actually want to attach a new attribute to the existing table? A short answer is that such an operation for adding new (derived) attributes simply does not exist. We simply have no choice and must adopt what is available – a set operation.

*Link attributes.* Another generic data processing pattern consists in computing links (or references) between tables: given a record in one table, how can we access attributes of related records in another table? For example, assume that `Price` is an attribute of a second `Products` table (Fig. 1, right), and it does not exist as an attribute of the `Items` table. We have two tables, `Items` and `Products`, with attributes `ProductId` and `Id`, respectively, which relate their records. If now we want to compute the `Amount` for each item then the price needs to be retrieved from the second `Products` table. This task can be easily solved by copying the necessary attributes into a new table using the relational (left) join:

```sql
SELECT i.*, p.Price                           (2)
FROM Items i
JOIN Products p
ON i.ProductId = p.Id
```

This new result table has the necessary attributes `Quantity` and `Price` copied from two source tables and hence it can be used for computing the amount. Yet, let us again compare this solution with the problem formulation. Do we really need a new table? No. Our goal was to have a possibility to access attributes of the second `Products` table (while computing a new attribute in the first `Items` table). Hence, it again can be viewed as a workaround and forced solution where a new (unnecessary) table is produced just because it is the only way to access related data in this set-oriented model.

*Aggregated attributes.* The next typical data processing task is data aggregation. Assume that for each product in `Products`, we want to compute the total number of items ordered (Fig. 1). Group-by operation provides a standard solution:

```sql
SELECT ProductId, SUM(i.Quantity) AS TotalQ
FROM Items i
GROUP BY ProductId                            (3)
```

Again, we produce a new table although the real goal was adding a new (aggregated) attribute to the `Products` table. Our intention was to make `TotalQ` equivalent to all other attributes in the `Products` table so that it could be used for computing other product properties. Apparently, this also could be done in SQL but then we would have to apply join to combine the group-by result (3) with the original `Products` table to bring all attributes into one table like (2) followed by yet another set operation like (1) for calculating new attributes.

In all these examples, the problem formulation does not mention and does not actually require any new table. Yet, the applied data processing model provides only set operations, which means that it is a problem-solution mismatch. The necessity to adapt set operations for the task of defining and adding new attributes is not a problem of only SQL or the relational model (RM) [2]: it exists in all models and frameworks, which rely on set operations for data processing. In particular, we can see it in MapReduce [4] where map and reduce operations always produce new collections even if the goal is to compute a new object field or a new aggregated property, respectively. In this situation, there is no choice: we must use sets for all kinds of data operations even when they do not match the problem at hand and no sets need to be produced at all.

Adopting set operations for deriving new attributes has quite significant negative consequences at different levels of data organization. If multiple tables are being processed then we can easily get a conceptual mess: many different types of joins (inner, left, right, full), nested joins and intermediately computed and aggregated attributes in these tables all packed in one SQL statement.

### 1.2 What Is to Be Done?

A general solution to this problem consists in introducing a *column-oriented* data model providing operations for *directly* manipulating columns without changing the table data these columns belong to and hence doing precisely what is required: adding derived attributes to existing tables. Below, for illustration purposes, we show how such a hypothetical data model could be applied to the examples described before.

Adding a new calculated attribute could be done as follows (compare it to (1)):

```
CREATE ATTRIBUTE Amount
FROM Items TO Double // Mapping
AS Quantity * Price // Definition
```

A solution to the problem of accessing data in related tables is well known: it is based on introducing link or reference attributes, which can be then accessed using dot notion supported by queries. Such attributes contain values, which provide access to records in other tables. In our example, we need to define a new attribute in the `Items` table, which references records in the `Products` table (compare it to (2)):

```
CREATE ATTRIBUTE Product
FROM Items TO Products // Mapping
AS ProductId == Id // Definition
```

Now we can easily compute `Amount` using dot notion even though `Price` belongs to the `Products` table:

```
CREATE ATTRIBUTE Amount
FROM Items TO Double // Mapping
AS Quantity * Product.Price // Definition
```

What we have achieved here is simple and natural semantics of links: everybody understands what a link is and how to use it via dot notation. We also separated two different concerns - link definition and link usage – which are now in two statements by making it easier to maintain this code: if later we change how `Items` and `Products` are related then the way the amount is computed needs not be changed.

It is also possible to add an attribute, which will compute its values from subsets of records in another table. Such a query in our pseudo code could look as follows:

```
CREATE ATTRIBUTE TotalA
FROM Products TO Double // Mapping
AS SUM(Items.Amount) // How to aggregate
GROUP Items.Product // How to group
```

In contrast to calculate and link attributes described above, this attribute computes its values by aggregating data stored in another table. It relies on the previously defined link attribute `Product` for grouping but does not include its definition. Therefore, the grouping condition can be easily changed later independently without modifying other queries that use it.

These queries demonstrates the central idea behind our approach: we define a new attribute in an existing table instead of defining new unnecessary tables.

## 1.3 Contributions and Outline

Of course, there are tools, patterns and best practices, which can significantly help in writing such queries and data processing scripts, for instances, by translating them into SQL, MapReduce or another conventional set-oriented language. Yet, we argue that the demonstrated problem-solution mismatch is not a minor drawback, but rather a major problem caused by the application of wrong tools and the absence of right methods. Our goal therefore is not to fix conceptual problems of one layer by introducing yet another layer of complexity. It consists in finding a principled solution by developing a new data model, which can solve such tasks directly without the need to adapt inappropriate mechanisms.

In this paper, we describe a data model, called the *concept-oriented model* (COM), which is intended for representing and processing data using mathematical *functions* as opposed to using only sets and set operations in existing set-oriented data models and data processing frameworks. COM is able to manipulate functions as first-class elements. For example, `Amount`, `Product` and `TotalQ` in COM are (derived) functions and no new sets will be produced during inference.

COM radically changes the way we think of data by significantly strengthening the role of functions. In many (but not all) cases, it is possible to represent and process data by using only functions without changing the sets. In particular, COM has the following two important properties:

- Data can be stored in functions in the same way as it can be stored in sets. In particular, there can be two different databases, which have the same sets but different functions.
- Deriving (inferring) data in COM means computing new (mathematical) functions as opposed to producing new sets. In particular, a COM query may well produce a function (by processing data in other functions) rather than a set.

The idea of using functions for data modeling is not new and this branch has a long history of research starting from [6, 16]. COM can be viewed as a further development of the functional data modeling paradigm. Our main contribution in this context is that functions are made first-class elements of the *logical* model with the same status as sets. The existing functional models are either conceptual models (while COM is a logical data model) or heavily rely on set-oriented operations. They essentially extend the scope of a set-oriented model rather than providing a major alternative to set orientation. Conventional functional models emphasize that functions are important, should not be ignored and in many cases make data modeling easier (especially at conceptual level) but data management (at logical level) is still done mainly using sets and set operations like join and group-by.

Significantly strengthening the role of functions and making them first-class elements of the data model allows us to rethink the role of sets in data modeling. Formally, the role of sets is weakened because many tasks can be and should be solved by using functions. However, this weakening leads to significant simplification of the model as a whole. It can be viewed as a return to the original treatment of sets as collections of unique tuples by removing many complex and sometime controversial mechanisms arising from the necessity to use them for other purposes. In particular, COM does not need the following features: separation of relations and domains, the need in having primary keas and foreign keys, adopting set operations for performing calculations, aggregations and linking. All these mechanisms are now replaced by one formal construct, function, which makes the model simpler and more natural.

Such a simplification by reducing data modeling and processing to only two basic constructs - sets and functions treated in their original mathematical sense - would not be possible without rethinking some fundamental principles. In particular, we describe a functional alternative to describing data *connectivity*. COM assumes that two data elements are connected if there exists a function, which maps one of them to the other. It is opposed to the relational principle that data elements are connected if there exists a tuple in some set, which includes them as constituents. Obviously, these are two fundamentally different assumptions. Accordingly, COM assumes that connectivity is derived using *function composition* as opposed to *relation composition* (join).

The main general benefit of COM is that it does precisely what is requested: it allows us to define derived attributes without unnecessarily producing new tables.

In summary, our contributions are as follows:

- We argue that having only sets is not enough for data modeling and data processing and describe a new data model, which makes functions first-class elements of the model. Both sets and functions are equally used for data representation and data processing.
- We demonstrate how functions can represent the semantics of connectivity and how function composition can be used to derive new connections. This provides an alternative to the relational principles where relations are used for connectivity and relation composition (join) is used for inference.

- We describe how operations with functions can be used to solve some typical data processing tasks like computing new properties, data linking and data aggregation.
- We describe an open source framework, which is based on this data model and can be viewed as a functional alternative to MapReduce and other similar set-oriented languages and frameworks.

This paper focuses only on the *logical level* of data modeling and does not discuss any conceptual or physical aspects of data management. In particular, the column-orientation in the paper does not relate to column stores (even though the implementation uses columnar format for storing data). We describe the Bistro [1] toolkit only to illustrate one possible implementation of COM and do not discuss such (important) aspects as physical data organization, dependencies and topology of operations, incremental evaluation, optimization of function evaluation etc. This approach to data modeling and processing was also used for self-service data integration and analysis [11,12].

Note also that the above examples were provided using an SQL-like pseudo code to make it easier to comprehend the main motivation behind this research. The open source framework we describe is implemented differently and is closer to how MapReduce works where data processing logic is described programmatically as a graph of operations. In our code examples, we follow the convention that lower case identifiers like `product` denote (Java) objects while upper case identifiers like `Product` refer to (column and tables) names.

The paper is organized as follows. In Section 2, we introduce sets and describe how a purely set-oriented model can be used for data modeling by emphasizing the arising problems. In Section 3, we introduce functions and describe how they can be used for data modeling by solving the problems arising in a purely set-oriented approach. In Section 4, we describe operations with functions and demonstrate how they can be used for data processing. Section 5 describes how COM modifies set operations. Section 6 provides a summary, concluding remarks and outlook for future research.

## 2 Sets for Data Modeling

### 2.1 Sets and Values

In the Concept-Oriented Model (COM), the main unit of data is a *value*. Values can be only copied and there is no possibility to represent them indirectly via other values or share them. Examples of values are numbers like 45.67 or letters like 'b' represented using an appropriate encoding convention.

Any data value is supposed to have some structure. Values the structure of which is hidden or ignored are referred to as *primitive values*. Values with an explicitly declared structure are referred to as *complex values*. Complex values are made up of the copies of other values and this composition is formally represented by a *tuple*, which is treated in its accepted mathematical sense by capturing the notion of an ordered list.

A tuple consisting of *n* member values is called *n*-tuple and *n* is called *arity*. Tuple members are enclosed in angle brackets and their position is referred to as an *attribute*. Attribute names are separated from the member values by a colon. For example, $e = \langle a: x, b: y \rangle$ is a complex value composed of two values *x* and *y* having attributes *a* and *b*, respectively. The values *x* and *y* might also have some structure. One value can be part of many different tuples in the form of multiple copies. It is not permitted to include a value into itself.

The *empty tuple* $\langle \rangle$ without any structure is treated as a special data element denoted as NULL. We assume that adding empty value to or removing it from a tuple (independent of its position) does not change the tuple: $\langle a: x, b: y, c: \langle \rangle \rangle = \langle a: x, b: y \rangle$.

A collection of unique values is formally represented as a mathematical *set*. Sets capture the very simple notion of a group or collection of things. Importantly, a set is a collection of distinct tuples and hence no element can appear more than once in the same set. For example, $S = \{x, y, z\}$ is a set consisting of three values *x*, *y* and *z* which must be distinct. We will assume that any value is a member of some set and it is possible to determine the set a value belongs to. The notation $x \in S$ is used to denote that the value *x* is a member of the set *S*.

The *empty set*, written as {} or $\emptyset$, is a special set which does not contain any values. In mathematics, it is also assumed that $\forall S, \emptyset \subseteq S$.

### 2.2 Set Membership for Data Modeling

A class of set-oriented data models rely on only sets of tuples for modeling data. Accordingly, a generic set-oriented *database* is defined as a number of sets each consisting of some tuples (complex values):

$D = \{S_1, S_2, \dots\}$

Here $S_i = \{v_{i1}, v_{i2}, \dots\}$ are sets, and $v_{ij} = \langle u, w, \dots \rangle$ are tuples composed of values from other sets.

*Structural constraints*. In general, tuples within one set may have any attributes which differ from tuple to tuple. It is possible to impose structural constraints by specifying a list of attributes and their *types* which are allowed for the set:

$S = \{\langle v_1, v_2, \dots \rangle\}$, where $v_i \in S_i$

---
[1] https://github.com/asavinov - Bistro data processing toolkit

Now tuples may have only certain structure by including only values from the specified sets called types.

Below we enumerate some properties of set-oriented models:

- *Set nesting*. According to this definition, a set-oriented model does not support nested sets, that is, a set consists of only tuples and cannot include other sets as its members. Although such a support could be theoretically provided, this feature makes the model much more complicated. At the same time set nesting can be modeled using flat sets and references.
- *Two kinds of sets*. Many concrete set-oriented models introduce two kinds of sets. For example, RM distinguishes between domains and relations, which are both normal sets but play very different roles in the model. Other models like the functional data model (FDM) distinguish between value sets and entity sets. Although practically having two kinds of sets can be useful, it is a controversial decision from the theoretical points of view. Indeed, a set is a set and there have to be a really strong reason to introduce different kinds of sets.
- *Tuple nesting*. Nested tuples are naturally supported by this category of models because a tuple member can be a tuple with its own tuple members up to primitive values.
- *Flattening nested structure*. In some models like RM, the nested structure of tuples is flattened by removing intermediate levels so that any tuple consists of primitive values. It is a highly controversial feature from the data modeling point of view because we essentially discard important information about the structure.
- *Inclusion by-value*. Both nested and flat tuples support only inclusion by-value. This means that a tuple consists of *copies* of its member values. In particular, there is no possibility to reference elements in other sets or include them using other indirect ways.

Probably the most important property of set-oriented models is that the only basic relationship is set membership: either a (data) value $x$ is a member of a set, $x \in S$, or not, $x \notin S$. The only way to change a set-oriented database (at the basic level) is either *adding* a tuple to a set or *removing* a tuple from a set. Importantly, no other operations are supported. Why it is a significant limitation is discussed in the following subsections describing more specific aspects of data modeling.

## 2.3 Modeling Identifiers. Primary Keys and Surrogates

Although adding things to and removing them from a collection is a very general modeling pattern, the question is what the thing is? In this context, there exists another important modeling pattern (not only in data modeling): things we model are typically uniquely identified, which means that there exists something unique that can be "detached" from it and then used to access it. In data modeling, this detachable part is frequently referred to as an *identifier* and can be implemented as a pointer, reference, surrogate, link, primary key or a similar construct.

What is in an identifier, their roles and uses is big topic but we would like to emphasize only two their benefits:

- They significantly decrease the amount of (redundant) data being transferred and stored because only some (small) part of the represented thing is copied.
- The represented thing can be modified without the need to update all its numerous copies (the identifier itself is supposed to be immutable). It is essentially a mechanism of sharing data.

An important observation is that a purely set-oriented model does not support the mechanism of identifiers. We can only manipulate a whole thing by adding it to a set or removing it. The whole thing in this case is supposed to identify itself and can be stored or transferred only by-copy. For example, if we want to represent an order item then we create a relation with attributes characterizing this order item including quantity ordered, price and date. If we want to represent this order item in some other set then we must copy the whole tuple including all its attributes. There is no other choice if we do not want to modify the underlying theory and fundamental properties of sets and tuples.

Obviously, such a model is extremely inconvenient and there exist several general solutions. One approach is based on introducing an additional layer on top of the base set-oriented model, which can be characterized as a subset of attributes used for identification. In RM, such a subset is called a *primary key* (PK). However, the mechanism of PKs has one fundamental flaw. Tuple as a whole becomes a mutable data element. For example, we can change the quantity of an order item, and this change does not produce a new thing – we still have the same order item because it has the same PK. Thus, PKs change the fundamental principle of set-orientation: tuples are immutable and can be only added or removed. Apparently, the cause of the problem is that we still assume that a thing (identifier and properties) is represented by one tuple, that is, we follow the principle "one tuple – one thing".

There exist also other problems with PKs like the controversy with the treatment of inclusion. On one hand, we want to include only PK in other tuples and hence only PK is treated as a true tuple. On the other hand, we still treat all attributes as a tuple because it is how a relation is defined. The controversy is that we cannot unambiguously answer the question whether a set consists of PKs only or a set consists of whole things (PK and non-PK attributes).

Another solution consists in introducing some built-in identifiers typically implemented as surrogates [5], oids [7], references or system identifiers. It is somewhat similar to PKs because we break all attributes into two groups. The main and important difference from PKs is that surrogates are managed by the system and hence are not part of the model (and not part of the tuples). This has some benefits and drawback. An advantage is that surrogates are immutable (while PK typically can be changed) and have many other properties of true identifiers implemented by the system. A significant drawback is that it is not possible to define their domain-specific structure (in contrast, PKs may have arbitrary user-defined structure). Yet, from the fundamental point of view, we still have the controversy: does a set consist of surrogates (as its tuples) or it consists of whole things (surrogate and properties)? If a set consists of only surrogates then only surrogates can be processed, which is useless in most cases (because we want to process data in properties). If we assume that a set consists of whole things (surrogates and properties) then we break the fundamentals of set-orientation because tuples become mutable and we must copy them into other tuples.

Why do we want to answer these questions and resolve the controversies? Because we want to use *formal* set operations for data processing rather than rely on specific properties of ad-hoc mechanisms and additional layers. These controversies can be resolved by introducing functions and we describe this in Section 3.3.

## 2.4 Modeling Properties. Foreign Keys and References

In the previous subsection, we emphasized the importance of having identifiers and inability to support them without sacrificing some major principles of set-orientation. In this section, we discuss how we can model thing properties, that is, the other side of identifiers. Assume that we know an identifier of a thing (e.g., modeled by PK or surrogate). The main question now is how we can use it to *access* properties of the represented thing?

Accessing a property normally means two operations: *getting* a value stored in the property and *setting* (assigning) a new value to the property (by overwriting the old one). Here we see a fundamental difference of this data manipulation pattern from the add-remove pattern. Indeed, we do not want to add or remove anything – we are thinking about something existing and want to simply modify it. A pure set-oriented model does not support such a pattern but there exist workarounds, which simulate it using set operations or some other mechanisms and assumptions.

One wide spread approach to implementing the update operation is based on the mechanism of foreign keys (FK). Here the idea is that some values stored in the attributes of this tuple are associated with the values stored in attributes of another relation. This allows us to find a tuple in another relation given values stored in this relation. Normally it is assumed that only PK is stored in other relations. Formally, FK is a constraint, which allows for using only values already existing in the target relation. The idea is that attributes from two related tables are copied into one table by using the relational join operation, by matching tuples from the source relations. The main problem of this approach [8] is that FKs have the semantics of references and properties while the operations provided along with the mechanism of FK are set-oriented (Section 1.1).

Another approach is based on built-in system identifiers (surrogates, oids, references etc.) so that associations between values in different sets are maintained by the system. It could be viewed as an ideal solution because the system supports dot notation in queries and we do not have to think how to read and write values of properties. However, it is too far from this status for one reason: this mechanism of access cannot be customized because it is not part of the model. Essentially, it is the same problem as we have with references and surrogates. We solve the problem for the price of losing control over how things are identified and how things are accessed.

Thus, the choice is either to have full control over identification and access by using rather inappropriate and complex set operations, or first-class support of access operations without control over its implementation. COM solves this problem by satisfying both of these requirements and in Section 3.4 we describe how this mechanism based on functions works.

## 2.5 Modeling Objects

Things can be modeled by representing them as objects or entities [1]. Objects are different from and opposed to values and it is a fundamental observation. Values are passed by copying their constituents while objects are passed by-reference and hence can be used for sharing data. Since it is a wide spread data modeling pattern, the question is how objects can be represented using sets and tuples? For example, how a product (object) can be thought of as and formally represented via tuples in sets?

Many generalizations of set-oriented models [17, 3] make a principled assumption that tuple attributes represent fields of one object and hence a set stores a number of objects of the same class (in COM it is not so). This approach suffers from one controversy: mathematically, a tuple is a value passed by-copy while an object is not a value because it is passed by-reference. In order to resolve it, we need to mark these (entity) attributes as having a special status by essentially excluding them from the tuple. However, if object fields do not belong to the set tuples then how they should be treated formally? Obviously, it is analogous to the controversy arising due to the introduction of PKs and FKs.

## 2.6 Modeling Connectivity. Joins

One fundamental question is how different tuples are related and what does it mean for tuples to be connected? RM provides a clear answer:

> $n$ values $v_1, \ldots, v_n$ are (directly) related if there exists a tuple $\langle \ldots, v_1, \ldots, v_n, \ldots \rangle \in S$ where they are members (in any order and possibly combined with other values)

This is why the set $S$ is referred to as a *relation* in RM – its tuples relate values from the domains. If we want to connect some existing values then the only way is to create a new set and add a tuple, which is made up of the related values. This type of connectivity is symmetric, that is, all values have the same status. This connectivity relationship is also *n*-ary, that is, 2 or more values can be related. One serious restriction of RM is that only values from domains can be related – tuples from arbitrary relations cannot be explicitly related because tuple attributes cannot contain other tuples (due to flattening). Yet, relations between arbitrary tuples can be modeled indirectly by including all their attributes (which is highly unnatural).

This definition allows us to model direct connections between values. In order to infer indirect connections, we need another assumption, which defines what is meant by inference. RM uses *relation composition* for deriving a new relation given two input relations. If $R \leq X \times Y$ and $S \leq Y \times Z$ are two (binary) relations then their composition $S \circ R$ is a set of $\langle x, z \rangle$ pairs:

$$S \circ R = \{\langle x, z \rangle \in X \times Z | \exists y \in Y: \langle x, y \rangle \in R \land \langle y, z \rangle \in S\}$$

The idea is that initially (before inference) two values $x$ and $z$ are not included into any tuple and hence they are not directly related. However, (different) tuples they are included into contain one common value $y$ and hence they are indirectly related. This idea of inference is based on the property of including some common value. In other words, if two values have some common parts then they are related. Note that this operation is also symmetric and it allows for indirectly connecting more than two values.

This semantics of connectivity has inherently set-oriented nature and provides a very powerful formalism for inferring new sets from existing sets by essentially adopting the principles of the logic of predicates. The question however is how relevant this semantics of connectivity and mechanism of inference is for data modeling? It is proven to be useful for many use cases and data modeling patterns. However, as we demonstrated in the introduction, there exist quite general scenarios, which do not use one tuple as a representation of a relation between values. In addition, the relational connectivity semantics does not directly model the concept of objects where we distinguish between an identifier and properties. Although relation composition (join) is used for accessing properties given an identifier in the FK pattern, this support is not very natural because we apply a set-oriented pattern (derive a new set) without having such a need (as described in Section 1.1). Therefore, we will describe new function-oriented semantics of connectivity in Section 3.6 and show how it is used to model and derive connections.

## 3 Functions for Data Modeling

### 3.1 Functions and Value Mappings

Mathematically, a function is a mapping from a set of input values into a set of output values where exactly one output is associated with each input: $f(x): D \to R$. Here $D$ is a set of all input values, called *domain*, $R$ is a set of all output values, called *range*, and $x$ is an argument that takes its values from $D$. There are two conventions for representing an output given input: $y = f(x)$ and $y = x.f$ (dot notation).

A function can be represented as a set of input-output pairs: $f = \{\langle x, f(x) \rangle, x \in D\}$. One pair in this set is referred to as a *function element* and (like any tuple) it is a value. This representation is useful for formal reasoning but since it hides the semantics of functions (as a mapping), we will not use it. For data modeling, we assume that

- a set is a collection and hence we can add or remove its member values
- a function is a mapping and hence we can get or set its output values

In the case some input has no output value explicitly assigned it is supposed to be NULL (empty tuple). Therefore, all inputs have exactly one output assigned. Yet, for the purposes of this paper, we assume that functions take only non-NULL values.

### 3.2 Value Mappings for Data Modeling

A concept-oriented *database* is defined as a number of sets and a number of functions between these sets:

$$D = \langle S, F \rangle$$

where $S = \{S_1, S_2, \ldots\}$ are sets, $F = \{f_1, f_2, \ldots\}$ are functions, $f_i: S_j \to S_k$, $i = 1, 2, \ldots$, for some $j$ and $k$. It belongs to a category of function-oriented models because function is an explicit element of this model used for data representation and (as we show later in the paper) data processing. It is a generic definition and depending on the constraints imposed on the structure of functions and sets, we can get more specific types of data models. For example, we could prohibit cycles of functions or we could introduce an (unstructured) model with only one set (the universe of discourse) and functions representing mappings between its elements. However, studying such (important) cases is not the purpose of this paper.

In this definition, it is important that functions have the same status as sets but different semantics and purpose. Data representation and data processing is not limited by sets only. In addition to sets, we can represent data using functions and process data by producing new functions from other functions. In particular, two databases may have identical sets but different functions and hence they are different databases (which is not possible in purely set-oriented data models).

A database *schema* is a database without set elements and function elements. To define a schema, it is necessary to specify its sets (without their members) as well as functions along with their input and output sets but without function elements. If a schema has been defined then it is treated as a constraint, which means that, set members and function members must obey this structure.

Note that in this definition, functions are distinguished from tuple attributes. A value stored in an attribute is part of the tuple where the attribute is defined. In contrast, outputs of functions are not stored in the input tuple. (In [10] we referred to attributes and functions as identity functions and entity functions, respectively.)

How functions are used to represent data? In contrast to sets where the basic operation is adding and removing tuples, the basic operation with functions is *getting* a value and *setting* a value for a given input tuple. Setting a value is essentially *assignment* operation and it is precisely what is absent in the set-oriented paradigm. Note that assigning an output value to some input value does not change any set, that is, we can manipulate the state of a database without changing set membership relation. Manipulating data in a function-oriented database means changing the mappings between its sets where the sets represent existing things.

Since the state of the functions (mapping) has to be stored somewhere we say that functions are viewed as a data store along with sets. Note that storing a function is not directly related to the column store technology. In column stores, we assume that all data is represented as a table and the question is whether to *physically* represent it as a row store or column store. In COM, the task is to physically represent functions independent of the sets.

Another possible confusion comes from the formal possibility to represent a function as a *set* of input-output pairs. This suggests that there is actually no need to introduce a dedicated construct – function – we can model everything using sets. Here it is important to understand that sets and functions have different semantics, and this difference is of crucial importance for data modeling. In other words, function membership can be and should be used only for formal analysis or for physical representation but not as its semantics. In data modeling, we treat functions as *mappings* and can only get or set their outputs.

### 3.3 Modeling Identifiers via Tuples

In a set-oriented model, it is a controversial issue (Section 2.3). By introducing functions, this controversy is resolved. Now we have an unambiguous answer: any tuple within a set is an identifier for something. A set is then a collection of identifiers. Tuples in sets have the semantics of existence (no properties or characterization). If a tuple is added to a set then it represents a thing, which is supposed to really exist, and if it is removed from the set then this thing is supposed to be non-existing. The main benefit is that there is no need in having such mechanisms as PKs or surrogates. The semantic load on sets is significantly reduced and the whole model gets simpler. More details about semantic differences between identities and entities can be found in [15].

### 3.4 Modeling Properties via Functions

Legalizing functions as first-class elements of data models essentially means that we recognize that mappings between things are as important as things themselves. Moreover, things without mappings represent a formally degenerated and practically quite useless model. This view contrasts with the purely set-oriented paradigm where the complete data state is represented and all operations are performed by using only sets.

In COM, tuples have only one main usage: they manifest the fact of existence of a thing, which essentially means that a tuple is an identifier of a thing. How then such things are characterized? This is done by means of functions. Namely, a function is treated as a property and its output is treated as a value of this property. Thus, the primary purpose of functions is characterizing things using other things.

The usage of functions for characterizing things has the following important features:
- Data is manipulated by getting and setting function outputs as opposed to adding and deleting tuples in the case of set operations
- Functions differ from attributes because changing a function output does not change any set while changing an attribute changes the set. Functions allow us to characterize things without changing the thing identifier.
- Properties and functions essentially turn values into references and introduce the mechanism of data access by-reference. In other words, a reference is a normal value, which can be used to retrieve other values using functions
- Properties and functions provide a mechanism of sharing data. If we change some property (by setting a new function output value) then all other elements storing this input will see the new value without the need to update them
- Function names become an important part of the data model because they need to be specified in data processing scripts

## 3.5 Modeling Objects

Functions allows us to simply and naturally solve the problem of representing entities (or objects). In COM, an object is a number of function output values returned for the same input value *e* which is treated as the object identifier or reference:

$$E(e) = (f_1(e), f_2(e), \dots)$$

Here we used round brackets in order to distinguish it from tuples denoted by angle brackets. An object always has some identifier, which is an (input) tuple in some set. Object fields are also values but they are stored in arbitrary sets. Importantly, an object is not a tuple. In particular, it is not possible to pass one object as whole in one operation because these values are stored separately and are available only by using the corresponding functions.

For example, a product can be identified by its number and hence product numbers are values within the set of products. A product object, however, is defined by its properties, which are represented by functions, and the function output values are stored in other sets. More specifically, if a product object is characterized by its name and price then these two functions map each product number into some values in the corresponding (string and numeric) primitive sets.

## 3.6 Modeling Connectivity

COM uses semantics of connectivity, which is based on functions:

> two values *x* and *y* are related or connected if there exists a function *f* which maps value *x* to value *y*

In other words, two tuples are related if one of them is mapped to the other one by means of some function. If we need to relate two sets then it is done by defining a new function between them. The main distinguishing feature of this approach is that the way data values are related is determined by functions. In particular, we can change the way elements in the database are connected without changing its sets. In contrast, the set-oriented approach assumes that connections between elements are determined by sets and hence we need to modify some set in order to change connections between elements.

Functions provide a direct way to connect values. New connections can be derived using *function composition*. This operation combines two or more consecutive mappings into one mapping by applying the next function to the output of the previous function. Formally, if $f: X \to Y$ and $g: Y \to Z$ are two functions, then their composition is a new function: $h(x) = g(f(x))$

Alternatively, function composition can be written using dot notion, $g(f(x)) = x.f.g$ , or circle notion, $g(f(x)) = (g \circ f)(x)$. This way to derive new data is a functional counter part of relation composition (Section 2.6). The main difference is that instead of producing new sets it produces new functions.

# 4 Functions for Data Processing

## 4.1 Manipulating Data Using Functions

The currently dominating approach to manipulating data is based on set-oriented principles where deriving new data means defining a new set with tuples composed of tuples from already existing sets. In this section, we describe a function-oriented approach where deriving new data means defining a new function using already existing functions.

Such function-oriented data processing is based on two basic operations:
- Getting function output given some input: $y = f(x)$ or $y = x.f$
- Setting function output for some input: $f(x) = y$ or $x.f = y$

These operations are widely adopted in programming but they are not suitable for data processing where we want to manipulate *collections* of objects rather than individual objects. What is worse, these operations do not reflect the semantics of typical data processing patterns: it is not specified what it means to read or assign a function output. Therefore, main questions are how such basic functional operations can be used to solve typical data processing tasks.

We consider three general tasks, which can be solved by using functional operations instead of set operations. They correspond to the motivating examples in Section 1:
- *Computing* new function outputs directly from other functions in the same set (Section 4.2). Such definitions are referred to as *calculate functions* and they replace SELECT and Map set operations.
- *Finding* new function outputs using outputs of existing functions as criteria (Section 4.3). Such definitions are referred to as *link functions*. They are intended for linking sets and replace JOIN operation.
- *Updating* new function output values (multiple times for one input) using functions in another (related) set (Section 4.4). Such definitions are referred to as accumulate functions. They are intended for data aggregation and replace GROUP BY and Reduce.

It is important that we do not want to define functions by specifying *explicitly* their output values for all inputs. The way a function is defined should avoid iterators and loops over the input values. Therefore, functions will be defined by providing a mechanism of computing *one* output while it is the task of the system to apply this logic to all necessary inputs. For any definition type, we define a new function in an existing table using some other functions and the difference is only how its output values are computed: by computing, finding or updating values.

## 4.2 Calculating Function Output

Let us assume that new function outputs depend on only this table function outputs and the new output values can be *directly* computed from them. Such a function definition is referred to as a *calculate function* and it needs only one *expression*, which specifies how one output value is computed given other function outputs for one input. Given some value $x$ of the input set $X$ with functions $f_1, \ldots, f_n$, a calculate function output is represented by an expression:

$$f(x) = calculate(f_1(x), \ldots, f_n(x)) \in Y$$

Here *calculate* is an expression returning a value from $Y$ given outputs of the functions $f_1, \ldots, f_n$. Note that this expression processes individual values – not sets. This expression has to guarantee that the computed value really exists in the output set.

For example, let us assume that we have already a set `Items` with two functions `Price` and `Quantity`, which map each item to some numbers. Now we want to define a new function `Amount`, which computes the product of `Price` and `Quantity`. First, we create a column (function) object:

```
Column amount = db.createColumn(
  "Amount", // Column name
  items, // Input table
  objects // Output table
);
```

and then we provide a definition:

```
amount.calculate(
  x -> (double)x[0] * (double)x[1], // Lambda
  price, quantity // List of parameters
);
```

The first argument is a lambda expression, which returns the product of two parameters passed as an array. The second and third arguments are column objects this function depends on. The system evaluates this column by iterating through all elements of the `Items` set, retrieving the outputs of the functions `Price` and `Quantity`, calling the lambda expression by passing these two values as an array, and storing the expression return value as the `Amount` function output for the current input.

This approach solves the first problem we described in Section 1.1 by relying on only functions defined using a value-based expression computing its output directly from inputs and without any awareness of the sets existing in the model. It essentially is a functional analogue of the SELECT and Map operations but without the necessity to define and generate sets.

## 4.3 Finding Function Output

There exist data processing patterns where it is not possible to directly compute outputs of a new function. However, we can *find* an output element by using criteria expressed in terms of its properties. More specifically, given an input value $x \in X$ of a new function $f: X \to Y$, the output value $y \in Y$ is found by imposing constraints expressed as a predicate $p$:

$$f(x) \in \{y | p(x.f_1, \ldots, x.f_n, y.g_1, \ldots, y.g_m) = true\} \subseteq Y$$

This predicate connects $n$ properties of input $x$ and $m$ properties of output $y$.

The simplest and most useful predicate is equality, which means that we search for a tuple $y \in Y$ with properties equal to some properties of the input $x$:

$$f(x) \in \{y | y.g_1 = x.f_1, \ldots, y.g_n = x.f_n\} \subseteq Y$$

Although there can be many elements $y$ satisfying the predicate for one input, we will assume that either there is only one element or there exist additional criteria for choosing only one.

Finding an output satisfying certain criteria is formally based on the operation of inverting a function or de-projecting a value. An *inverse function* $\tilde{f}: Y \to P(X)$ returns a subset of inputs, which all map to the same output:

$$\tilde{f}(y) = \{x \in X | f(x) = y \in Y\}$$

We also can use inverse arrows '←' to denote the same operation:

$$\tilde{f}(y) = y \leftarrow f$$

Inverse arrow is opposite to dot notation and we use it [13, 14] because dot symbol does not have an inversion.

In the case we have many properties specified as a criterion, the operation of finding an output of a function is written as follows:

$$f(x) = y \in \bar{Y} = \bigcap_{i=1}^{n} Y_i = \bigcap_{i=1}^{n} \tilde{g}_i(z_i)$$

The function takes a value from the intersection of the de-projections of input value $x$ properties.

In practice, the way a system performs de-projection and finds an element satisfying certain criteria depends on the implementation, and there exist numerous techniques for optimizing such a search. At logical level, it is important only that we can define a new function by saying that its outputs have to be equal to certain input properties.

For example, if we have two isolated sets `Items` and `Products` (for example, loaded from CSV files) then we might need to define a function, which maps each order item to the corresponding product. First, we create a new column object by specifying its name, input table and output table objects:

```
Column product = db.createColumn(
  "Product", // Column name
  items, // Input table
  products // Output table
);
```

Now we can provide a definition for this column:

```
product.link(
  new Column[] { productsId } // In Products
  new Column[] { itemsProductId } // In Items
);
```

The first argument in this definition is a list of the output table properties: in this case only one column object `productsId` representing column `Id` in the `Products` table. The second argument lists the corresponding input element properties: in this example, only one column `itemsProductId` representing column `ProductId` in the `Items` table. For each input element from `Items`, the system will find an element from `Products`, which has the same id. It will then store them as outputs of this new column. After evaluation, this column can be used in other expressions to access products directly from order items.

This approach solves the second problem we described in Section 1.1 by relying on only functions defined by specifying search criteria. It can be treated is a functional analogue of the JOIN operation but without the necessity to define and generate sets. Although the way such functions are defined is very similar to join criteria, they are semantically completely different because here we define a function (mapping) rather than a set [8].

### 4.4 Updating (Accumulating) Function Output

Both calculate and link functions return a single final value of the function by directly computing it or by finding it, respectively. Importantly, an output depends on only one input tuple (and its properties). There exist a very important data processing task, called aggregation, which cannot be solved by using these functions because its result depends on *many* tuples, which are somehow related to the input and are referred to as a group. In order to compute an output value, the function has to process all tuples in the group. For example, assume that we want to compute the total sales of all products by defining a function, which maps each product to some number. Obviously, it does not depend on the properties of the product – it depends on line items stored in another table (but related to this product). This data processing pattern actually involves two separate tasks: grouping and aggregation. Both of these tasks can be solved by using only functions.

Grouping is performed using the following interpretation of a function. If $g(z): Z \to X$ is a function then de-projection $Z' = \tilde{g}(x)$ is a subset of tuples from $Z$, which are related to the element $x \in X$. Elements from $Z'$ are frequently referred to as *facts*, and elements from $X$ are referred to as *groups*. Function $g(z)$, called *grouping function*, assigns a group $x$ to each fact $z$ and, on the other hand, the inverse function $\tilde{g}(x)$ returns a subset of facts a group $x$ consists of.

In our example, `Items` contains facts and `Products` contains groups. The `Product` link column we defined in the previous section is a grouping function, which assigns a product to each line item.

Now let us consider how aggregation is performed. The task is to define a new function $f(x): X \to Y$, which computes its output $y \in Y$ by processing a subset $Z' = \tilde{g}(x)$. We could pass a subset of tuples $Z'$ to an aggregate expression, which will process them in a loop and return one value. However, it is precisely what we want to avoid because it breaks the whole conception by requiring an explicit loop and explicitly processing subsets. The problem can be solved by introducing *accumulate functions*, which get only one fact $z \in Z'$ as well as some output value $y \in Y$:

$$y' = update\big(y, f_1(z), \dots, f_n(z)\big) \in Y$$

The task of this function is to *modify y* by using $n$ properties of the fact $z$ and return the updated $y'$ result. This update expression is completely unaware of the loops and groups – it processes individual values. The idea of aggregation using such accumulate expressions is that the system calls it for each fact by passing the previous return value as an input for the next call:

$y^0 = C \in Y$ where $C$ is an initial value
$y^j = update\big(y^{j-1}, f_1(z^j), \dots, f_n(z^j)\big) \in Y, j = 1, \dots, k$
$x \leftarrow g = \tilde{g}(x) = Z' = \{z^1, \dots, z^k\}$
$f(x) = y^k$

The initial value is some constant like 0. Then we update this value using properties of the fact $z^1$ and get a new output $y^1$, which is then used to call again the update expression but with the properties of the next fact $z^2$ and so on. The last value $y^k$ will be the final value of the function $f$ being evaluated for the input $x$. It is necessary to call the update expression $k$ times for $k$ facts $z^1, \dots, z^k$ from the group in order to compute the output for one input $x$.

In order to compute total sales for each product in our example, we create a new column by specifying also a default value:

```
Column total = db.createColumn(
  "Total", // Column name
  product, // Input table
  objects // Output table
);
total.setDefaultValue(0.0);
```

and define it using an accumulate expression:

```
total.accumulate(
   product, // Grouping (link) column
   (a,x) -> (double)a + (double)x[0], // Lambda
   amount // Fact properties to be aggregated
);
```

The first argument of this definition is a reference to the grouping column (defined in the previous sub-section as a link column), which maps items to products. The second argument is a lambda expression, which adds the amount of the fact (it is a calculate column) to the current (intermediate) aggregate value for the product. It will be called as many times as there are order items for this product. The third argument is a reference to a property of the items, which is being aggregated.

Note that this definition uses two derived columns – product (grouping function) and amount (aggregated property) – but for defining new functions it is not important and it is one of the benefits of this approach because we can define and, more important, later modify various properties independently.

This approach solves the third problem described in the introduction by providing a function-oriented replacement for such set operations as GROUP-BY or Reduce. The main benefit of accumulate functions is that no new unnecessary sets are produced and it relies on only normal value-based expressions requiring no loops or iterations [9].

## 5 Sets for Data Processing

### 5.1 Manipulating Data Using Sets

In the previous section we described how new data can be derived by defining new functions, and it was assumed that the sets are not changed during inference. Although calculate, link and accumulate functions can replace some general set-oriented data processing patterns, there still exist some cases where it is necessary to derive a new set and not a function. We consider three such tasks:

- *Product* of several sets
- *Filtering* a set
- *Projecting* a set

The main difference of all these operations from their set-oriented analogues is that new sets are defined in terms of functions, and they produce new functions as their result in addition to a set. These new functions connect the result set to the source set(s) and hence the result set is not isolated. We can always use these new functions to access other sets and their functions in other definitions. It is important because there is no need to copy all the original data into each new result set – they can be accessed from the result using the connections (functions) between sets.

### 5.2 Combining Tuples – Product of Sets

One important data modeling and data processing pattern consists in finding all combinations of tuples in two or more existing sets. This operation is one of the corner stones of multidimensional analysis because the source sets can be treated as axes (with tuples as coordinates) and the product set treated as a multidimensional space (combinations of coordinates representing points).

COM supports the product operation, which is defined as follows:

$$Y = X_1 \times ... \times X_n =$$
$$= \{\langle a_1: x_1, ..., a_n: x_n \rangle | x_i \in X_i\}$$

Although formally it is the conventional Cartesian product, it has the following distinguishing features:

- In contrast to RM, the result set is not *flattened* and each result tuple has *n* attributes each being equal to some tuple from a range set.
- The product set is a *derived* set with the population automatically inferred from the existing sets. Thus, COM can infer both functions and sets.
- The product set retains its *connection* with the source sets. We can always access source tuples given an output tuple using attribute names what is useful when defining other sets or functions.
- The product reflects the semantics of *multidimensionality* and the operation is not intended for expressing connectivity via joins as it is in RM. Although formally we can use it for joining (by adding some filtering conditions), semantically it will mean that we are using RM and not COM.

For example, let us assume that we have two tables: Products with a list of products and Quarters with a list of quarters (like 2018Q1. 2018Q2 etc.) For multidimensional analysis, we might need to build a table (cube) of all their combinations. First, we create a table object representing a multidimensional space, and add two columns, which will represent the corresponding axes:

```
Table pq = schema.createTable("PQ");

Column product =
   db.createColumn( "Product", pq, products );

Column quarters =
   db.createColumn( "Quarters", pq, quarters );
```

Second, we define this table as a product of two other tables:

```
pq.product();
```

After such a definition, this table will be automatically populated by all combinations of product and quarter records, that is, each tuple in this table is a cell identified by one product and one quarter.

The next step would be adding new derived functions characterizing the cells and it can be done as described in Section 4.4 using accumulate columns, which use facts from the `Items` table.

### 5.3 Filtering Tuples

Filtering records is one of the most widely used operations and its purpose is to select records from a table, which satisfy certain conditions. COM allows for filtering records using the product operation rather than a dedicated operation:

$$Y = X_1 \times \ldots \times X_n =$$
$$= \{\langle a_1: x_1, \ldots, a_n: x_n \rangle | p(x_1, \ldots, x_n) = true\}$$

If we now leave only one source dimension in the product then it will be a filter:

$$Y = \{\langle a: x \rangle | p(x) = true\}$$

Essentially, this means that the result set $Y$ will contain records from $X$, which satisfy the specified predicate. The filtered table will have one attribute which points to selected records from the source table.

For example, we could select all products with low prices:

```
Table cheap = schema.createTable("Cheap");
Column product =
  db.createColumn("Product", cheap, products);
cheap.product(
  x -> (double)x[0] < 100.0, // Lambda
  new ColumnPath(product, price)
);
```

First, we create a table object, which will store filtered records. Second, we create a column, which points to a table with source records. Finally, we provide a definition with the first argument being a predicate lambda expression returning true if the parameter is less than 100.0. The second arguments specifies a parameter. In our example, it is a sequence of two column segments (represented by the `ColumnPath` class). The first column segment starts from the new table and leads to the source table `Products`, and the second segment retrieves the price of this product.

### 5.4 Projecting a Set

Assume that there is only table `Items` and no table `Products` but the task is to compute various properties of products like sales amount. In this case, we simply do not have a table to attach these properties to. A list of products could be restored by enumerating all *unique* product identifiers occurring in the `Items` table. This set operation is called *projection* along a function. It is applied to a source set $X$ by specifying one of its functions $f$ and results in a new set $Y$, which consists of all unique outputs of this function:

$$Y = X \to f = \{f(x) | x \in X\}$$

The arrow here is analogous to dot in dot notation with the difference that it is applied to sets (and also it allows for inverting this operation).

In our system, project columns are used for projection, and they are defined in the same way as link columns with one difference: link columns do not change the output set while project columns will automatically populate it.

## 6 Conclusion

The main motivation for this research is based on the observation that applying exclusively sets and set operations is inappropriate for many wide spread use cases because they are actually aimed at deriving new columns rather than tables. Since existing models and data processing frameworks provide mainly set operations, this leads to the need to define multiple tables without necessity. This makes data models and data processing scripts more complicated, difficult to write, comprehend and maintain.

As a general solution, we described a new data model, called the concept-oriented model (COM), which relies on both sets and functions as two primary data modeling constructs. In comparison to purely set-oriented models (like RM), COM significantly reduces the semantic load on sets by treating them in their original mathematical sense as collections of tuples and only collections of tuples without any additional mechanisms and assumptions like PKs, FKs, domains vs. relations etc.

Functions are arbitrary mappings between sets. In comparison to existing functional models, their semantic load increases:

- functions represent properties (instead of FKs)
- functions represent connectivity (instead of joins)
- functions allow us to introduce objects (as combinations of their outputs)
- function provide a mechanism of access by-reference and dot notion
- functions represent a portion of the state of the database so that two databases with the same sets could differ by their functions
- functions are used for inference by deriving new functions from existing functions
- functions are used for linking
- functions are used for aggregation

We described how COM can be used for data processing by introducing three functional operations, calculate, link and accumulate, as well as some set operations. We also described one possible implementation of this approach in an open source toolkit intended for general purpose data processing and designed as a functional alternative to MapReduce.

The main benefit of introducing functions as first-class elements is that models as well as data processing scripts become simpler, more natural, easier to design and maintain because the data modeling constructs provided by COM (functions and operations with functions) do precisely what is necessary in many use cases – directly defining a new column.

There are several directions for future research:

- semantic and conceptual aspects of COM including inheritance, polymorphism, semantic relationships, multidimensional models, NULL values
- expanding this approach on other data processing use cases like stream processing and big data processing
- architecture and system design aspects including topology organization (a graph of set and function operations), dependency management, incremental evaluation (propagating small changes through the topology), performance issues